# Code Vulnerability Detection Across Different Programming Languages with AI Models

Hael Abdulhakim Ali Humran
Department of Artificial Intelligence and Data Science
Istanbul Aydin University
Istanbul, Turkey
habdulhakimhumran@stu.aydin.edu.tr

Prof. Dr. Ferdi Sönmez
Department of Artificial Intelligence and Data Science
Istanbul Aydin University
Istanbul, Turkey
ferdisonmez@aydin.edu.tr

*Abstract*— Security vulnerabilities present in a code that has been written in diverse programming languages are among the most critical yet complicated aspects of source code to detect. Static analysis tools based on rule-based patterns usually do not work well at detecting the context-dependent bugs and lead to high false positive rates. Recent developments in artificial intelligence, specifically the use of transformer-based models like CodeBERT and CodeLlama, provide light to this problem, as they show potential in finding such flaws better. This paper presents the implementations of these models on various datasets of code vulnerability, showing how off-the-shelf models can successfully produce predictive capacity in models through dynamic fine-tuning of the models on vulnerable and safe code fragments.

The methodology comprises the gathering of the dataset, normalization of the language, fine-tuning of the model, and incorporation of ensemble learning and explainable AI. Experiments show that a well-trained CodeBERT can be as good as or even better than some existing static analyzers in terms of accuracy greater than 97%. Further study has indicated that although language models can achieve close-to-perfect recall, the precision can decrease. A solution to this is given by hybrid models and validation procedures, which will reduce false positives.

According to the results, the AI-based solutions generalize to different programming languages and classes of vulnerability. Nevertheless, robustness, interpretability, and deployment readiness are still being developed. The results illustrate the probabilities that AI will enhance the trustworthiness in the usability and scalability of machine-learning-based detectors of vulnerabilities.

## I. INTRODUCTION

The contemporary software systems are very complex and quite frequently developed in a plurality of programming languages, which elevates their vulnerability surface considerably. Security deficiencies have to be detected as early as possible so as to minimize the risks that may arise in the development process. By conventional wisdom, the most common approach to locating the vulnerabilities would be through Static Application Security Testing (SAST) tools, which are essentially rule-based methods, including the concept of pattern matching, symbolic execution, and data flow analysis. Such tools as Mythril and Slither on smart contracts or general-purpose analyzers in C/C++, Java, and Python have proven to work in intimate settings. Nevertheless, these tools have heavy false positive rates, a scarce ability to identify sophisticated vulnerabilities or those not covered before, and poor quality of work domains under multilingual circumstances owing to the existence of syntactic and semantic diversities [1], [2].

The recent development of artificial intelligence has been seen especially in the branch of deep learning models that use a transformer, which provides an opportunity to develop more intelligent and flexible approaches to vulnerability detection. Large and diverse models (such as CodeBERT and CodeLlama) can be trained to consider complex uses of syntax and semantics within code and interesting or novel code connections. These models are credible because they frame the problem of vulnerability detection as one based on natural language processing (NLP) and thus show the potential to surface more subtle bugs that are otherwise overlooked by other methods [3].

This research endeavor is inspired by the fact that hard-coded rule-based systems are not suitable in areas where speedy developing heterogeneous software environments are required. The AI models are able to generalize programming languages and learn emerging coding patterns rather than following a preknown signature of a vulnerability like a static analyzer. This flexibility is of particular concern to large-scale systems created in more than one language and where requirements reform fast. Transformers such as CodeBERT do not only increase the accuracy of detection, but they also decrease false alerts and offer better generalization in the case of vulnerabilities that they were not specifically trained on.

The paper is research on transformer-based models towards detecting cross-language vulnerabilities. It aims at refining CodeBERT and CodeLlama with multilingual vulnerability-ready sets and measures their accuracy, recall, and generalization potential. Ensemble techniques and explainable AI are also incorporated in this research to maximize the interpretability of the model and the false positive rate. Moreover, they are compared to the classic SAST tools to evaluate how they apply in practice. The contribution made by this work is related to the fact that the focus toward performance, trust, and practical user-friendliness helps create viable and applicable AI-powered solutions to easily implement secure software development.

## II. LITERATURE REVIEW

Recent studies have shown that deep learning has been very effective in identifying vulnerabilities in programs. The first approaches relied on RNNs and LSTMs to treat source code as sequences, whereas graph-based models used the structure of a program, including ASTs and PDGs [3]. These methods gave birth to transformer-based models, including CodeBERT and GraphCodeBERT, which are trained on millions of pieces of code and can be used as feature extractors due to their strong capabilities [4]. The relatively promising performance of the recently emerged Large Language Models (LLMs), such as CodeLlama and LLaMA 3, on vulnerability classification after the fine-tuning on the datasets, such as Big-Vul and DiverseVul, exceeded 95% F1-scores on a binary classification [3].

DetectBERT is a line-level vulnerability detector proposed by [5] that does not use any graph input. It focuses on robustness and generalization using real-world data, including CVEFixes and VUDENC, and also on code normalization procedures. DetectBERT performed well in the detection of Python vulnerability and exceeded the performance of conventional graph-based models.

The multilingual models like CodeBERT can learn across languages and types of vulnerabilities [2]. There are alternative solutions, and one of them is proposed by [6], who mentioned the method of creating vulnerability databases through the mining of actual security patches in GitHub repositories of Python. By doing so, they produced a high-quality dataset that allowed successful fine-tuning of the model, such as Qwen-7B, with the potential that the transfer learning to similar languages would work.

In the sphere of smart contracts, the so-called SmartLLM proposed by [2] combines the fine-tuned LLaMA model and Retrieval-Augmented Generation (RAG) to understand the context. The model got an accuracy of 100% on the recall but with lower precision, thus making it ideal in critical fields where accepting false negatives is not an option.

Wang proposed the PFSCV that combines CodeBERT and UnixCoder to join semantic and structural code characteristics. This method achieved more than 94 percent sensitivity on vulnerability types of smart contracts, such as the reentrancy [4].

To mitigate the problem of false alerts, [1] provided VVF-AI, or a two-stage system, which has an AI-based agent that checks vulnerabilities identified by the first checker. The agent tries to simulate or reason about exploitability reaching 93.1 % verification rate of various forms of vulnerability.

Steenhoek performed a user study DeepVulGuard, which is an IDE plugin based on CodeBERT and GPT-4. For developers, the tool was accurate, and it provided interactive feedback but has a limitation in the context capabilities and false positive. This is the same reason why tools AI must be on par with developer workflows [7].

Approaches that are based on anomalies consider the vulnerabilities as deviations in a normal behavior. Li introduced the initial methodology to utilize an anomaly-attention transformer based on the labels of the execution traces generated by fuzzing [8]. The accuracy of their model was 87.7 percent in finding the vulnerabilities in binary programs compared to the CNN-based and LSTM-based models [9].

Das examined the question of whether they are based on real vulnerability characteristics or artifacts. They demonstrated that their assessment based on perturbations suggests that many models produce false positives on patched code, spurring the need to train with better methods and explainable models [10].

Maturi included visualization based on attention on top of a BiLSTM model to show the lines of code that had the most impact on the predictions. XAI techniques provide transparency and enable trust in model choices by the developers [9].
.

## III. METHODOLOGY

### a) Overview

The flow of vulnerability detection proposed has the main stages (1) preprocessing multilingual sets, (2) representation of code and tokenization, (3) selection and fine-tuning of a model (and ensembles), (4) evaluation with standard metrics, and (5) integration with explainability tools. The essential functionality is to classify the source code as vulnerable or safe in binary fashion and it can optionally be extended to report types of vulnerabilities.

### b) Dataset Preparation

In order to have a variety, we have listed code snippets in more than one language.

In the case of C/C++, we leverage Big-Vul dataset and more CVE samples that were considered as some of the most frequent issues, i.e., buffer overflow (CWE-119), integer overflow, and use-after-free (Curto et al. 2024).

In the case of Python, we mimicked the outline of CVEFixes and VUDENC, where we extract pairs of vulnerable-patched code, which encompass web-related vulnerabilities such as SQL injection and path traversal [5]. There was data normalization to minimize overfitting to identifiers.

In the case of Solidity smart contracts, we added SmartBugs and Ethereum CVEs of reentrancy, arithmetic overflows, and access control. To improve context, we incorporated Ethereum metadata that is done similarly in SmartLLM [2].

We did not expose our data to the risk of data leakage; therefore, we maintained a high level of separation between the training set and test set on the project level. The data has been divided into 80/10/10 training, validation, and testing, respectively. As seen in table 1, the distribution of the sample is summarized.

### c) Code Representation and Tokenization

CodeBERT was used to tokenize code based on multi-language support of its byte-pair encoding tokenizer that was optimized to operate on source code [11]. Context within which it was necessary was marked by tags given in language (e.g., <SOL>). Eight-bit (bytes) token sequences were maxed at 512, with long functions chopped up as required. We used line level annotation practices to pass line-sensitive planktonic information to the model [3].

### d) Model Architecture and Training

We pre-trained CodeBERT using a two-way classifier with cross entropy loss and label smoothing and weighting of the classes to support imbalanced data. Recall was enhanced with oversampling of vulnerable samples as well CodeBERT performed at high levels with ~97.00 percent accuracy and F1-score on the test data set.

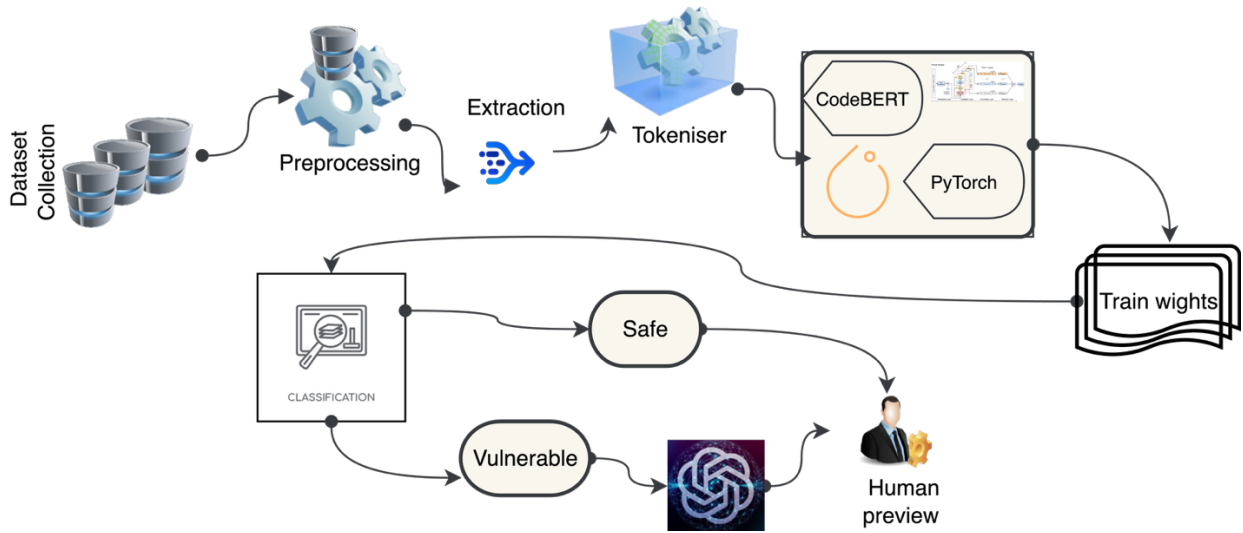

*Figure 1* Overall Architecture Pipeline of The Model.

To look at the architectural options, we have finetuned CodeLlama-7B in the LoRA way and developed a simple ensemble of CodeBERT and GraphCodeBERT, which has features of data flows [4]. The team did not recall much more but with decreased precision

The pipeline shown in Figure 1 incorporates CodeBERT as the basis encoder, a classification head, an optional ensemble with GraphCodeBERT, and post-checking in the shape of an explanatory and verification feedback loop based on GPT-4.

### e) Explainability and Feedback Loop

In order to have faith and complete inferences, we implemented an explainability module. In addition to VVF-AI [1] and DeepVulGuard [7], we solicited an interpretation of the models using GPT-4-based agents. Suspect or irregular predictions were either flagged and/or re-examined and it served to the clean-up process and explanation as well.

*Table 1 Dataset Summary*

| Dataset Name | Total Samples | Safe Samples | Vulnerable Samples |
|---|---|---|---|
| CVEFixes | 45,000 | 32,000 | 13,000 |
| Devign | 48,687 | 25,000 | 23,687 |

## IV. RESULTS

The proposed CodeBERT-based model was evaluated using standard classification metrics. It achieved 97.2% accuracy, with a precision of 98.05%, recall of 97.31%, and F1-score of 97.68%. These results reflect strong generalization and a low false positive rate (2.87%), outperforming many existing static and AI-based tools. For comparison, SmartLLM achieved perfect recall but only 70% precision [2], while DeepVulGuard achieved 80% precision but just 32% recall (Steenhoek et al. 2024).

**Figure 2** illustrates the loss convergence during training, while **Figure 3** presents the loss behavior during the final epoch. **Figure 4** shows the confusion matrix for the test set, highlighting accurate classification of vulnerable and safe code.

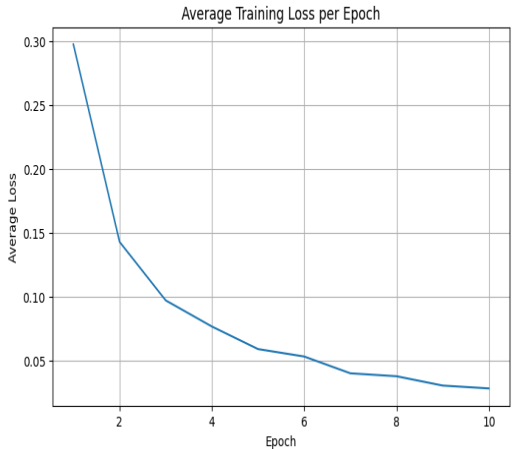
*Figure 4 Average Training Loss pe Epoch*

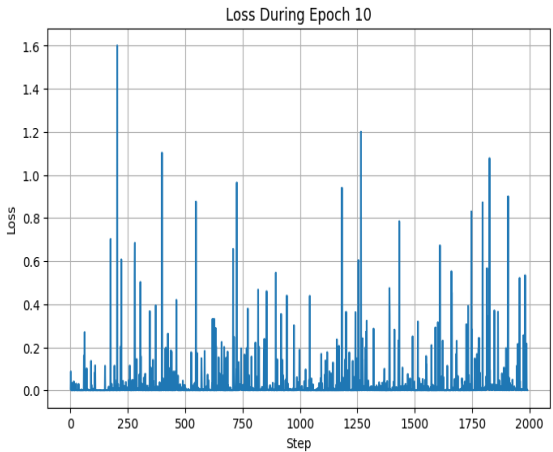
*Figure 3 Loss During Epoch 10*

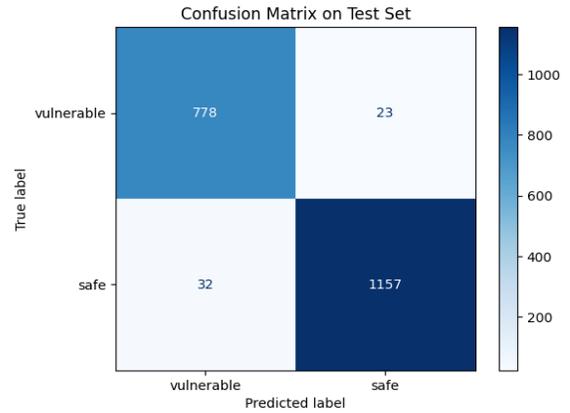
*Figure 2 Confusion Matrix on Test Set*

## V. CONCLUSION

This paper leads one to believe that the maturity of transformer-based ML models, including CodeBERT, has arrived at a point in which they are becoming powerful and, in some cases, even make teams with static analysis tools to detect vulnerabilities. They are strong in cross-language pattern matching, situational flexibility, and the capability to discover more complicated vulnerabilities than are restricted by handcrafted rules. Our model, which trains data diversity, has demonstrated high precision and recall, which decreased false negatives and, therefore, is a very useful tool in security-sensitive applications. Our demonstrations also revealed that integration of detection and verification by using LLM agents can be used to minimize false positives and develop developer credibility by imitating expert-like thinking.

Among the most notable contributions, it is possible to note the elaboration of a single multilingual model realized in different programming languages and showing generalizability even within the framework of uneven language presence. We combine explainability through attention modules and GPT-based rationalization, which is more transparent as well as allowing the use of these systems since they no longer present the machine-learning black-box concern.

Although encouraging, there are still issues to tackle concerning generality, the management of domain-specific or complex interprocedural vulnerabilities, scalability to low-resource settings, and robustness in the real-world flow of work. Its results correlate with the rest of the literature that implies that edge-based vulnerability detection is not just viable but also becoming increasingly feasible due to AI.

In the future, one of their studies into further research needs to be on ongoing learning, more robust verification methods, active learning by user suggestions, and hybrid AI involving both static and dynamic evaluation. The responsible deployment also ought to be directed to address the ethical concerns, such as dual-use risk and data privacy. The synthesis of empirical findings and the work of combining grounded empirical evidence and evidence that spurred new research forms a basis of creation of intelligent, reliable, and explainable security in current-day software development.

VI. REFERENCES

[1] C. Liu, T. Liu, Y. Tang, and J. Lin, "VVF-AI: A Vulnerability Verification Framework Based on AI-Agent," pp. 960–964, Jun. 2025, doi: 10.1109/AINIT65432.2025.11035850.

[2] J. Kevin and P. Yugopuspito, "SmartLLM: Smart Contract Auditing using Custom Generative AI," Feb. 2025, Accessed: Jul. 25, 2025. [Online]. Available: https://arxiv.org/pdf/2502.13167

[3] C. Curto, D. Giordano, D. G. Indelicato, and V. Patatu, "Can a Llama Be a Watchdog? Exploring Llama 3 and Code Llama for Static Application Security Testing," *Proceedings of the 2024 IEEE International Conference on Cyber Security and Resilience, CSR 2024*, pp. 395–400, 2024, doi: 10.1109/CSR61664.2024.10679444.

[4] D. Wang and S. Duan, "The smart contract vulnerability detection based on pre-trained model feature fusion," pp. 1761–1764, Jun. 2025, doi: 10.1109/ISCAIT64916.2025.11010394.

[5] S. S. Gujar, "DetectBERT: Code Vulnerability Detection," *2024 Global Conference on Communications and Information Technologies, GCCIT 2024*, 2024, doi: 10.1109/GCCIT63234.2024.10862235.

[6] K. Gladkikh and A. A. Zakharov, "Approach to Forming Vulnerability Datasets for Fine-Tuning AI Agents," *Proceedings - 2025 International Russian Smart Industry Conference, SmartIndustryCon 2025*, pp. 771–776, 2025, doi: 10.1109/SMARTINDUSTRYCON65166.2025.10986048.

[7] B. Steenhoek, K. Sivaraman, R. S. Gonzalez, Y. Mohylevskyy, R. Z. Moghaddam, and W. Le, "Closing the Gap: A User Study on the Real-world Usefulness of AI-powered Vulnerability Detection & Repair in the IDE," pp. 01–13, Dec. 2024, doi: 10.1109/icse55347.2025.00126.

[8] S. Li *et al.*, "Software Vulnerability Detection Based on Anomaly-Attention," *2022 4th International Conference on Robotics and Computer Vision, ICRCV 2022*, pp. 261–265, 2022, doi: 10.1109/ICRCV55858.2022.9953210.

[9] M. H. Maturi *et al.*, "Enhancing Smart Contract Security with Explainable AI: A Framework for Re-entrancy Vulnerability Detection and Explanation," *2025 IEEE Systems and Information Engineering Design Symposium, SIEDS 2025*, pp. 386–391, 2025, doi: 10.1109/SIEDS65500.2025.11021147.

[10] S. Das, S. T. Fabiha, S. Shafiq, and N. Medvidovic, "Are We Learning the Right Features? A Framework for Evaluating DL-Based Software Vulnerability Detection Solutions," pp. 2893–2904, May 2025, doi: 10.1109/ICSE55347.2025.00194.

[11] Z. Feng *et al.*, "CodeBERT: A Pre-Trained Model for Programming and Natural Languages," *Findings of the Association for Computational Linguistics Findings of ACL: EMNLP 2020*, pp. 1536–1547, Feb. 2020, doi: 10.18653/v1/2020.findings-emnlp.139.

**Please fill in the all authors' background:**

| Full Name | Email Address | Position | Research Interests | Personal Website (if any) |
|---|---|---|---|---|
| \multicolumn{5}{|c|}{**Position can be chosen from:**} |
| \multicolumn{5}{|c|}{**Prof. / Assoc. Prof. / Asst. Prof. / Lecture / Dr. / Ph. D Candidate / Postgraduate, etc.**} |
| Hael Abdulhakim Ali Humran | habdulhakimhumran@stu.aydin.edu.tr | Student | | |
| Prof. Dr. Ferdi Sönmez | ferdisonmez@aydin.edu.tr | Professor | | |